\documentclass[11pt, a4paper]{article} 
\usepackage{color}
\usepackage{upgreek}   
\usepackage{graphicx}
\usepackage{amsmath}
\usepackage{amssymb}

\usepackage{geometry} 
\usepackage[backend=biber, style=nature , doi=true]{biblatex} 
\begin{document}

\begin{center}
  
\textbf{\huge{Generation of photonic hooks from patchy microcylinders}}

\vspace{0.5cm}

\noindent {\large Fen Tang$^{1,2,\ddagger}$, Qingqing Shang$^{3,\ddagger}$, Songlin Yang$^{4}$, Ting Wang$^{5}$, Sorin Melinte$^{6}$, \\ Chao Zuo$^{7}$, Ran Ye$^{1,2,7}$*}

\end{center}

\noindent $^{1}$ School of Computer and Electronic Information, Nanjing Normal University, Nanjing, 210023, China\\
\noindent $^{2}$ School of Artificial Intelligence, Nanjing Normal University, Nanjing, 210023, China\\
\noindent $^{3}$ Key Laboratory for Opto-Electronic Technology of Jiangsu Province, Nanjing Normal University, Nanjing, 210023, China\\
\noindent $^{4}$ Advanced Photonics Center, Southeast University, Nanjing, 210096, China\\
\noindent $^{5}$ College of Materials and Chemistry \& Chemical Engineering, Chengdu University of Technology, Chengdu 610059, China\\
\noindent $^{6}$ Institute of Information and Communication Technologies, Electronics and Applied Mathematics, Universit\'e catholique de Louvain, Louvain-la-Neuve, 1348, Belgium \\
\noindent $^{7}$ Smart Computational Imaging Laboratory (SCILab), School of Electronic and Optical Engineering, Nanjing University of Science and Technology, Nanjing, 210094, China
\vspace{0.5cm}

\noindent \textit{*Correspondence: ran.ye@njnu.edu.cn}

\paragraph{Abstract:} The photonic hook (PH) is a new type of curved light beam, which has promising applications in various fields such as nanoparticle manipulation, super-resolution imaging, and so forth. Herein, we proposed a new approach of utilizing patchy microcylinders for the generation of PHs. Numerical simulation based on the finite-difference time-domain method was used to investigate the field distribution characteristics of the PHs. By rotating the patchy microcylinder, PHs with different curvatures can be effectively generated, and the PH with a bending angle of 28.4$^\circ$ and a full-width-half-maximum of 0.36 $\lambda$ can be obtained from 1 $\upmu$m-diameter patchy microcylinders.


\section{Introduction}
Photonic nanojets (PJs) are narrow, high-intensity, non-evanescent light beams generated at the shadow side of dielectric particles when irradiating the particles with light waves~\cite{04ChenZ}. They can propagate a distance longer than the wavelength of the incident light while keeping a minimum beamwidth smaller than $\lambda$/2, $\lambda$ is the wavelength of the incident light~\cite{09AlexanderH}. PJs show promising applications in various fields, such as micromaching~\cite{20YanB}, single-particle manipulation~\cite{16LiY}, optical sensing~\cite{15YangH, 18GuG}, super-resolution imaging~\cite{11WangZ, 13YeR, 16WangF, 16YangH}, and so forth. Within this context, many efforts have been made to design dielectric particles that can generate PJs with special characteristics~\cite{17BSLukYanchuk, 20MininIV3}.

In 2015, Minin et al. theoretically discovered a new type of PJ, which has a curved structure similar to the shape of a hook~\cite{15MininI}. They called the curved PJ as a photonic hook (PH). The PH is formed when electromagnetic waves pass through a dielectric trapezoid particle composed of a cuboid and a triangular prism. The combined effects of the phase velocity difference and the interference of the waves lead to a curved high-intensity focus~\cite{18YueL}. {The PH phenomenon was later experimentally observed in the terahertz and optical range~\cite{19MininIV, 20MininIV}. It can also be applied to the generation of curved surface plasmons~\cite{18MininIV}. }

PHs can be effectively generated from dielectric particles with asymmetric geometries, such as  dielectric trapezoids~\cite{18YueL, 19MininIV} and dielectric cylinders with glass cuboids inside~\cite{18YangJ}, or by using particles with an asymmetric distribution of refractive index (RI), such as Janus particles~\cite{19GuG}, Janus microbar~\cite{20GeintsYE}, and so forth. {Rotating the particles or adjusting the shape of the particles can effectively change the characteristics of the generated PHs~\cite{19GuG}.}
Special illumination conditions, such as partial illumination~\cite{20MininIV, 20LiuCY2} and nonuniform illumination~\cite{21GuG}, can also be used to generate PHs. {In this way, PHs can be generated using microcylinders with a symmetric geometry and a uniform RI distribution~\cite{20MininIV, 21GuG}.} In addition to obtaining PHs in the transmission mode, Liu et al. proposed the formation of PHs in the reflection mode~\cite{20LiuCY}, in which they used dielectric-coated concave hemicylindrical mirrors to bend the reflected light beams. {Geints et al. also proposed the formation of PHs in the specular-reflection mode under the oblique illumination of a super-contrast dielectric particle~\cite{21GeintsYE}.} Moreover, multiple PHs can be effectively generated using twin-ellipse microcylinders~\cite{20ShenX}, adjacent dielectric cylinders~\cite{20ZhouS} and two coherent illuminations~\cite{20ZhouS2}. {The PHs have promising applications in various fields, for example, nanoparticle manipulation and cell redistribution~\cite{19DholakiaK, 20MininIV3}.} Recently, Shang et al. reported the super-resolution imaging using patchy microspheres~\cite{21ShangQ}. Unlike conventional microspheres, which have a symmetric PJ, the patchy microspheres have a curved focusing and show an improved imaging performance due to the asymmetric illumination. Asymmetric illumination is a technique to enhance the imaging contrast in conventional bright-field microscopic systems~\cite{18SanchezC}, and now it is widely used in computational microscopic imaging to produce phase contrast~\cite{21FanY}. In addition, Minin et al. reported the contrast-enhanced terahertz microscopy under the near-field oblique subwavelength illumination based on the PHs formed by dielectric mesoscale particles~\cite{21MininOV}.

In this work, we show that the PHs can be generated using patchy particles of dielectric microcylinders that are partially covered with Ag thin films. Numerical simulation based on the finite-difference-time-domain (FDTD) method was performed to investigate the characteristics of the PHs. The spatial distribution of the Poynting vector and the streamlines of the energy flow in the simulated light field were given to illustrate the formation mechanism of the PHs. By adjusting the RI of the background, the diameter of the patchy microcylinder and the opening angle of the Ag films, PHs with various curvatures and intensity enhancement abilities can be effectively formed. In addition, the method of tuning PHs by rotating patchy microcylinders was also discussed in this paper.

\section{Simulation Method}

\begin{figure}[htb]
\centering
\includegraphics[width=0.8\textwidth]{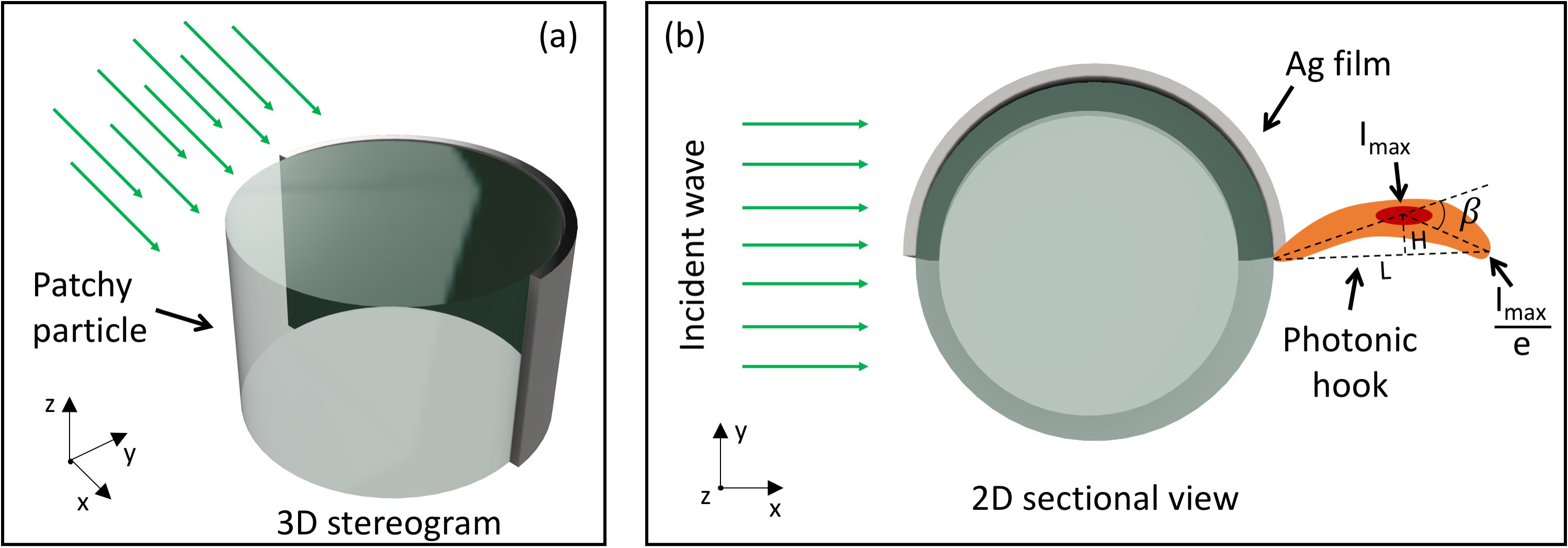}
\caption{Schematics of a patchy microcylinder illuminated by plane waves: (\textbf{a}) 3D stereogram and (\textbf{b}) 2D sectional view.} 
\label{Figure 1}
\end{figure}

Figures  \ref{Figure 1}a,b are the schematic drawing of the 3D stereogram and 2D sectional view of the investigated model. A dielectric microcylinder was created for two-dimensional simulation with the FDTD method {using Lumerical FDTD Solutions}. The top surface of the cylinder is covered with a 100 nm-thick Ag film. {As shown in Figure   \ref{Figure 1}b, an intense focusing of light will occur on the rear side of the cylinder when a P-polarized monochromatic plane wave ($\lambda$ = 550 nm) propagating parallelly to the X axis passes through the cylinder.} In this study, the RI of the cylinder is set to be 1.9, the same as the RI of BaTiO$_3$ (BTG), a high-index dielectric material widely used in microsphere-based applications~\cite{16WangF, 20YanB}. The diameter of the cylinder varies between 1--35 $\upmu$m and the RI of the background changes between 1.00--1.52. {For the entire computational domain, non-uniform meshes with RI-dependent element size were used and all of them are smaller than $\lambda$/50.} As shown in Figure  \ref{Figure 1}b, the PH’s degree of curvature is defined by the bending angle $\beta$, which is the angle between the two lines connecting the start point with the hot point, and the hot point with the end point of the PH, respectively. The hot point is defined as the point with the largest intensity enhancement (I$_{max}$), and the end point is defined as the point on the middle line of the PH with an intensity enhancement factor of I$_{max}$/e ~\cite{20MininIV, 20LiuCY}. {The hook height increment H and the subtense L of the curved photonic flux are also shown in Figure \ref{Figure 1}b.}

\section{Results and Discussion}

\begin{figure}[htb]
 \centering
\includegraphics[width=0.8\textwidth]{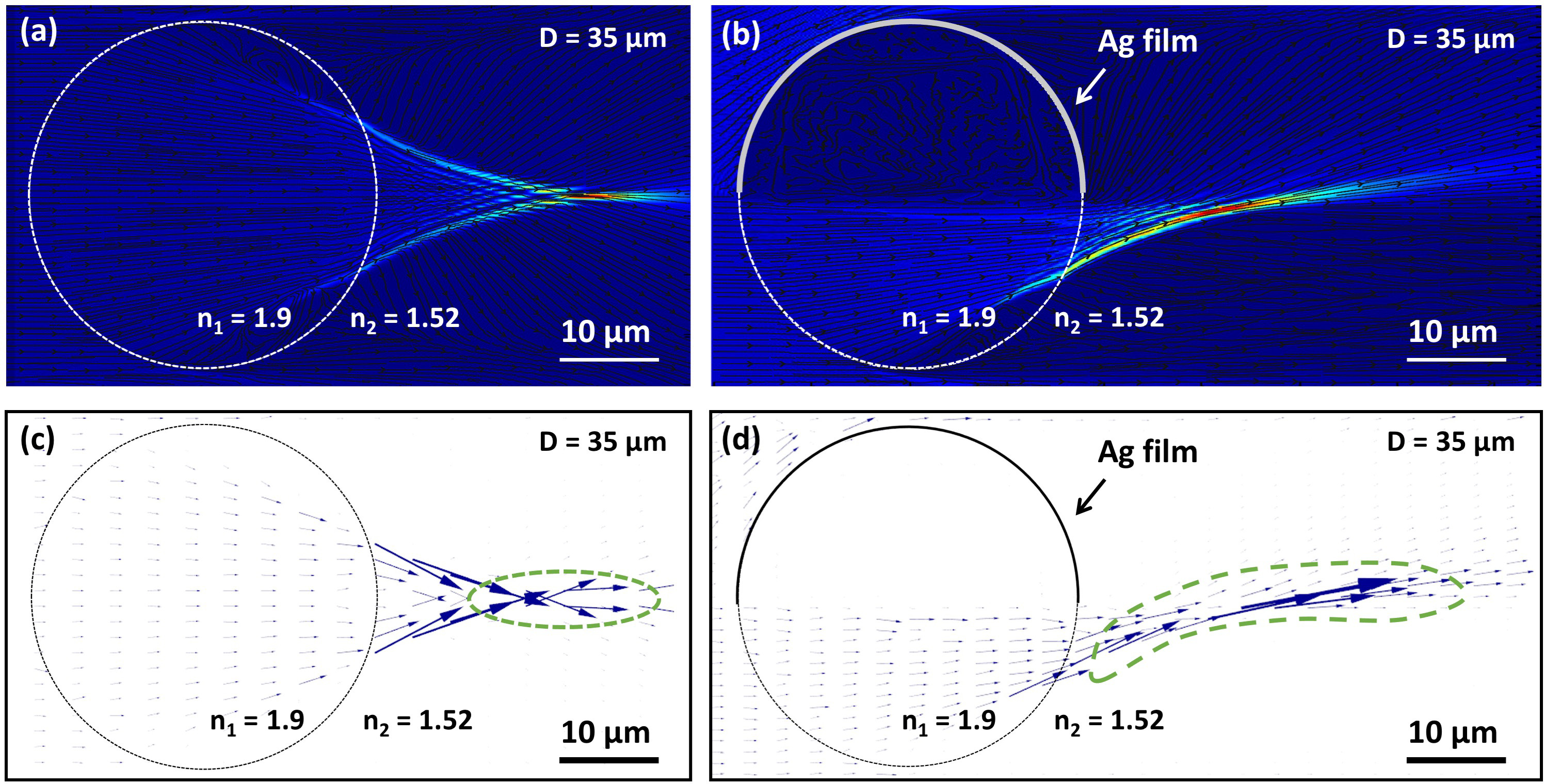}
\caption{(\textbf{a}) PJ formed by a 35 $\upmu$m-diameter pristine BTG microcylinder immersed in MIO; (\textbf{b}) PH formed by a 35 $\upmu$m-diameter patchy BTG microcylinder immersed in MIO; (\textbf{c},\textbf{d}) Corresponding Poynting vector of (\textbf{c}) the pristine microcylinder and (\textbf{d}) the patchy microcylinder.}
\label{Figure 2}
\end{figure}

First, we compared the optical field of the 35 $\upmu$m-diameter pristine cylinder and patchy cylinder under plane wave illumination. The background medium here is microscope immersion oil (MIO, n$_2$ = 1.52). As shown in Figure \ref{Figure 2}a, the incident light passing through the pristine cylinder forms a conventional PJ on the shadow side of the cylinder. The generated PJ has a symmetric $\mid$E$\mid$$^2$ distribution with the the midline of the PJ as the center of symmetry. On the contrary, as for the patchy cylinder shown in Figure \ref{Figure 2}b, the upper part of the incident light is blocked by the Ag film covered on the top surface of the cylinder, so only the lower part of the incident light can enter the patchy cylinder. {A curved light beam, that is, a PH, with a bending angle $\beta$ = 12.5$^\circ$, a hook height increment H = 1.51 $\upmu$m and a subtense L = 28.81 $\upmu$m is generated at the rear side of the cylinder.} Compared with the pristine cylinder, the patchy cylinder has a smaller intensity enhancement ability, as the I$_{max}$ is 36.0 and 14.3 for pristine and patchy cylinders, respectively.

As reported in the previous work~\cite{19GuG, 20LiuCY}, the formation mechanism of PJs and PHs can be analyzed using the time-averaged Poynting vector. In this work, the Poynting vector (blue conical arrows) of the optical field of the pristine and patchy cylinders under plane wave illumination is simulated with the FDTD method (Figure \ref{Figure 2}c,d), and the corresponding field-lines of the Poynting vector distribution are shown as the black lines in Figure \ref{Figure 2}a,b. As shown in Figure \ref{Figure 2}c, the spatial distribution of the Poynting vector inside and near the pristine cylinder is symmetric to the midline of the PJ (Figure \ref{Figure 2}c). Because the length of the conical arrows is proportional to the value of energy flux, the area containing longer arrows indicates a higher energy flux in that area. We can see that the energy flow corresponding to the pristine cylinder's optical field is focused into a classical PJ at the shadow side of the cylinder (Figure \ref{Figure 2}c). However, as for patchy cylinders, part of the incident light is reflected backwards to the space by the Ag film, which breaks the symmetry of illumination and makes the energy flow inside the microcylinder unbalanced. This asymmetric flow of energy is then focused into a curved beam after leaving the patchy cylinder, as shown in Figure \ref{Figure 2}d.

\begin{figure}[htb]
\centering
\includegraphics[width=0.8\textwidth]{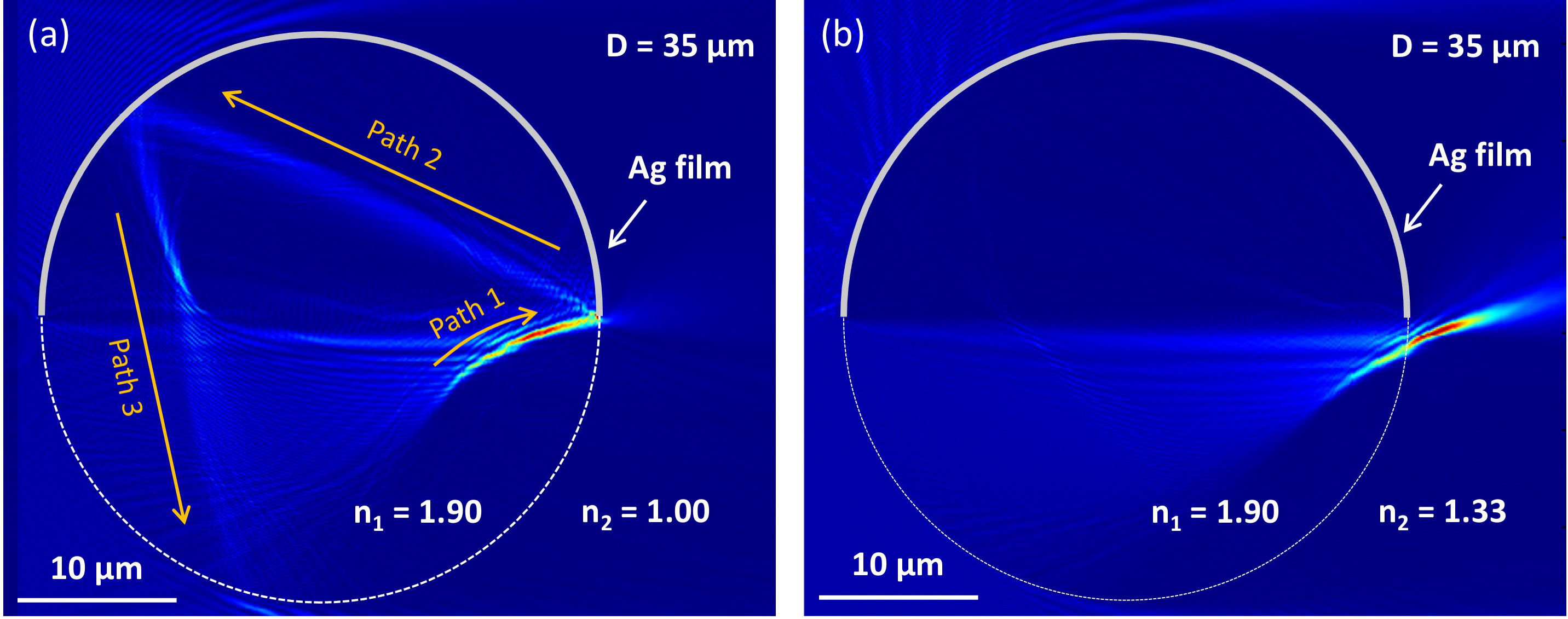}
\caption{Optical fields of 35 $\upmu$m-diameter patchy microcylinders immersed in (\textbf{a}) air and (\textbf{b}) water.}
\label{Figure 3}
\end{figure}

Next, the background medium was changed to air (n$_2$ = 1.0) (Figure  \ref{Figure 3}a) and water \mbox{(n$_2$ = 1.33)} (Figure   \ref{Figure 3}b) in order to investigate the influence of background RI on the characteristics of PHs. As shown in Figure   \ref{Figure 3}a, when the background RI is 1.0, the light entering the microcylinder will be reflected multiple times by the Ag film. The direction of light propagation is thus changed from Path 1 to Path 2 on the first reflection, and then from Path 2 to Path 3 on the second reflection. When the background RI is 1.33, the patchy cylinder forms a jet-like beam with a bending angle $\beta$ = 15.4$^\circ$. {The subtense and height increment of the beam are L = 6.17 $\upmu$m and H = 0.30 $\upmu$m, respectively}. The beam has the greatest intensity outside the cylinder with an enhancement factor of \mbox{I$_{max}$ = 24.8.} Considering that water is one of the most commonly used biocompatible materials, studying the PHs in water is of great importance for the practical applications of such curved beams. Therefore, we will use n$_2$ = 1.33 as the RI of the background medium in the following studies.

\begin{figure}[htb]
\centering
\includegraphics[width=0.8\textwidth]{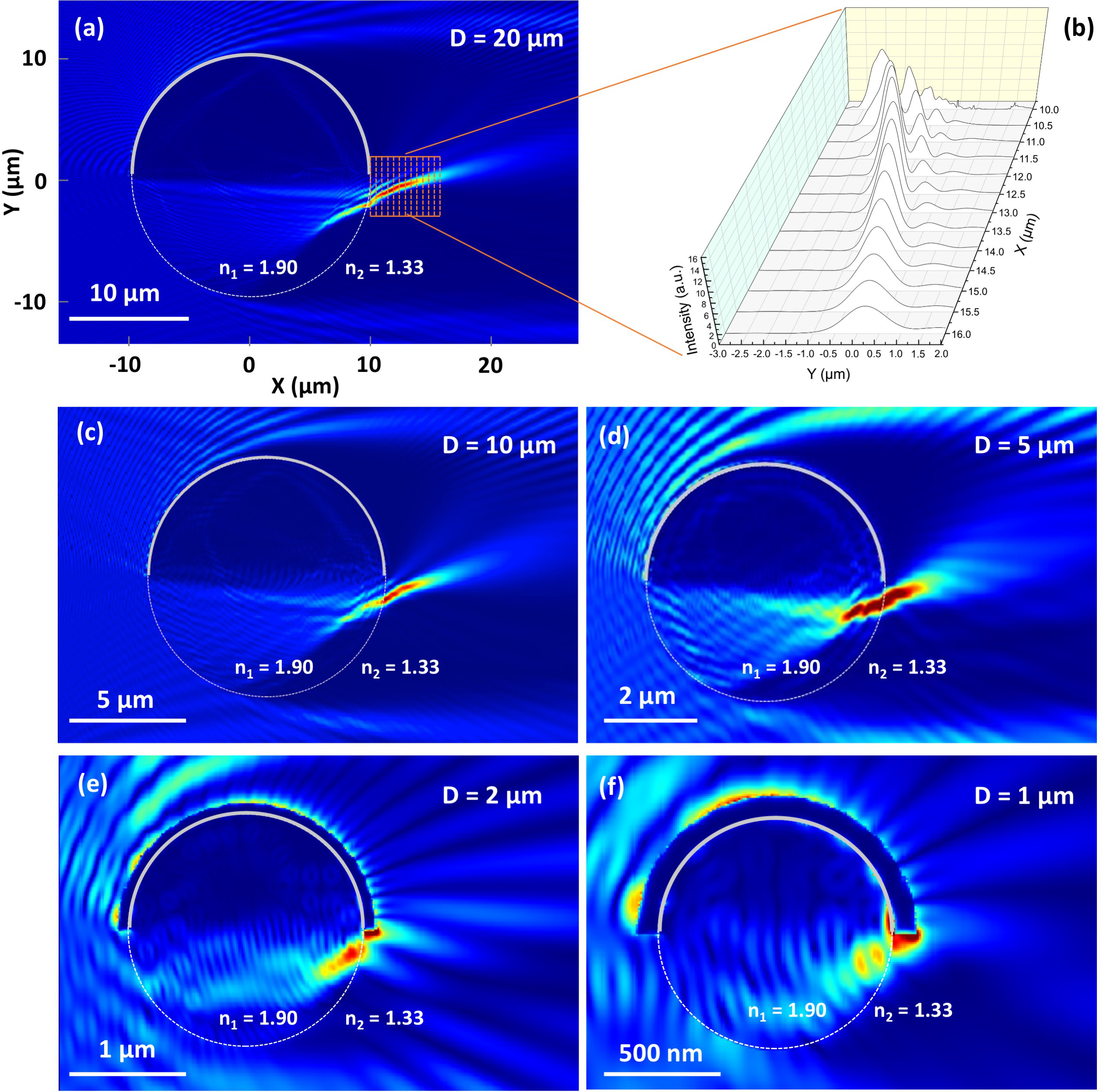}
\caption{(\textbf{a}) Optical field of the 20 $\upmu$m-diameter patchy microcylinder and (\textbf{b}) the cross-sectional profiles of the corresponding PH. (\textbf{c}-\textbf{f}) Optical fields of the patchy microcylinders with a diameter of (\textbf{c}) 10 $\upmu$m, (\textbf{d}) 5 $\upmu$m, (\textbf{e}) 2 $\upmu$m and (\textbf{f}) 1 $\upmu$m.}
\label{Figure 4} 
\end{figure}

The influence of particle diameter on the characteristics of PHs is also investigated in this study. Patchy cylinders of various diameters between 1 $\upmu$m and 20 $\upmu$m are illuminated with plane waves ($\lambda$ = 550 nm). The RI of the cylinder and the background medium is fixed at n$_1$ = 1.90 and n$_2$ = 1.33, respectively. {As shown in Figure \ref{Figure 4}a, a 20 $\upmu$m-diameter patchy cylinder generates a PH with a bending angle $\beta$ = 17.6$^\circ$, a hook height increment H = 0.52 $\upmu$m and a subtense L = 7.19 $\upmu$m.} The cross-sectional intensity profiles of the PH retrieved from the orange dash lines also confirm the curved trajectory of the I$_{max}$ position along the X axis (Figure \ref{Figure 4}b). {We found that the PHs generated by smaller patchy cylinders have a slightly greater curvature and a smaller subtense length and height increment, as the 10 $\upmu$m- and 5 $\upmu$m-diameter patchy cylinders can produce PHs with $\beta$ = 18.8$^\circ$, \mbox{L = 3.03 $\upmu$m,} \mbox{H = 0.10 $\upmu$m} (Figure \ref{Figure 4}c) and $\beta$ = 20.2$^\circ$, L = 1.33 $\upmu$m, H = 0.05 $\upmu$m (Figure \ref{Figure 4}d), respectively.} The intensity enhancement ability is found to be weaker in small particles. The \mbox{I$_{max}$ = 13.3} (Figure \ref{Figure 4}a), 10.4 (Figure \ref{Figure 4}c), 7.3 (Figure \ref{Figure 4}d) for 20 $\upmu$m-, 10 $\upmu$m-, 5 $\upmu$m-diameter patchy cylinders, respectively. This difference in intensity enhancement ability is due to the fact that large particles can collect more light waves and focus them onto a narrow space, leading to a higher I$_{max}$ at the focal point. However, there is no PH phenomenon in the optical field of patchy cylinders when the cylinder diameter is reduced to 2 $\upmu$m \mbox{(Figure \ref{Figure 4}e) }and \mbox{1 $\upmu$m} (Figure \ref{Figure 4}f), because the light scattering of dielectric particles with a diameter close to the wavelength of the incident light tend to be localized in the forward direction and no jet-like fields can be generated~\cite{17BSLukYanchuk}.

\begin{figure}[htb]
\centering
\includegraphics[width=0.8\textwidth]{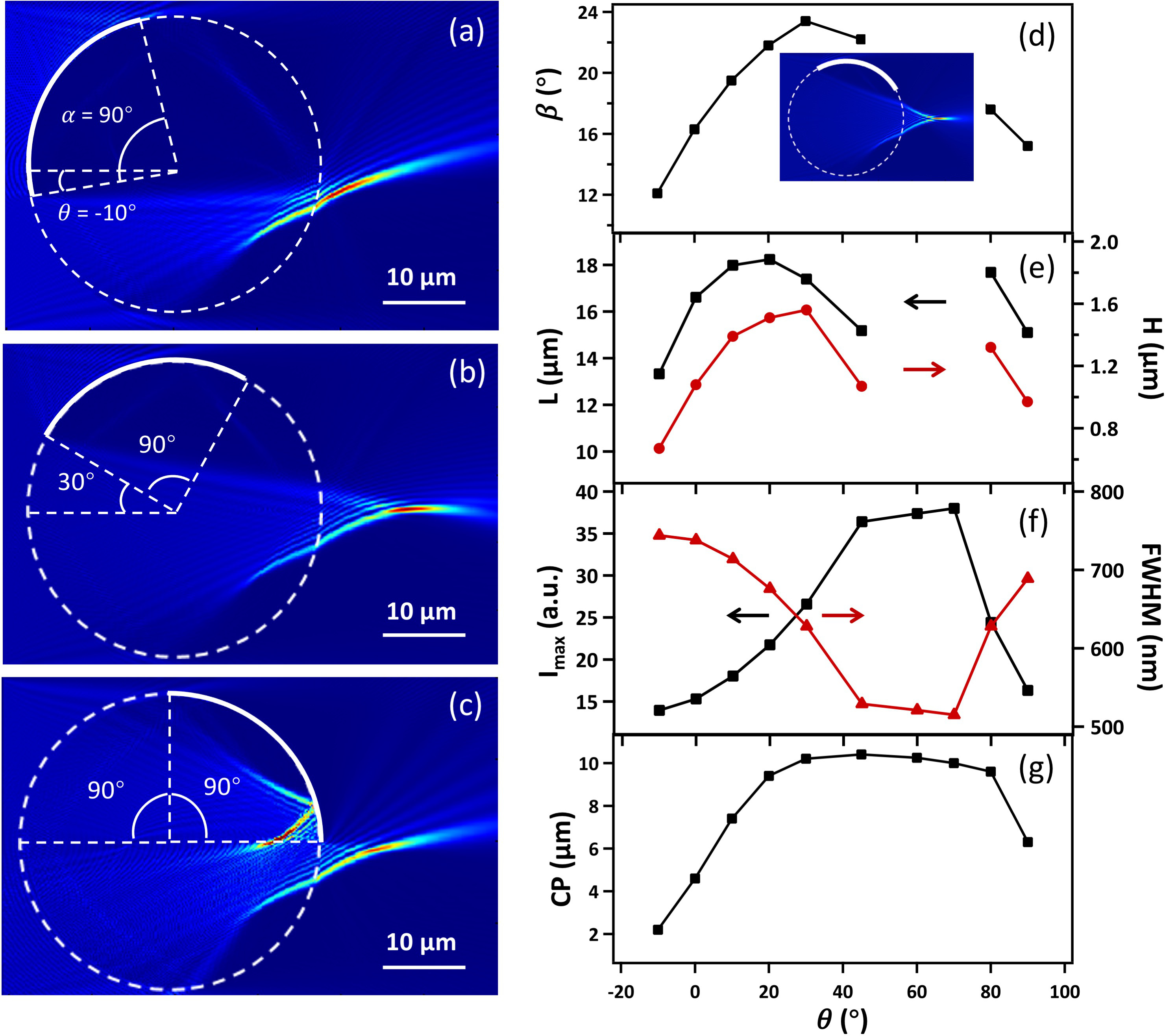} 
\caption{(\textbf{a}-\textbf{c}) PHs generated by the 35 $\upmu$m-diameter patchy cylinder at different rotation angles: (\textbf{a})~$\theta$ = -10$^\circ$, (\textbf{b}) $\theta$ = 30$^\circ$, (\textbf{c}) $\theta$ = 90$^\circ$. {(\textbf{d}-\textbf{g}) Characteristics of the PHs as a function of rotation angle $\theta$: (\textbf{d})~bending angle $\beta$ and the focus of a patchy particle at $\theta$ = 60$^\circ$ (the inset), (\textbf{e}) subtense L (black line) and hook height increment H (red line), (\textbf{f}) maximum intensity enhancement factor I$_{max}$ (black line) and the corresponding FWHM (red line), (\textbf{g}) distance between the I$_{max}$ position and the right edge of the cylinder (CP).}}
\label{Figure 5}
\end{figure}

In this study, we showed that the curvature of PHs can be changed by rotating the patchy cylinder. As shown in Figure \ref{Figure 5}a, a 35 $\upmu$m-diameter patchy cylinder (n$_1$ = 1.9) fully immersed in water (n$_2$ = 1.33) with 1/4 of its surface covered by Ag films ($\alpha$ = 90$^\circ$) is used for demonstration. The patchy cylinder is rotated clockwise around the center of the cylinder. The rotation angle ($\theta$) is defined as the angle between the horizontal line and the left edge of the Ag film. $\theta$ is negative when the left edge of the Ag film is lower than the horizontal line, and positive when the left edge of the Ag film is higher than the horizontal line. As shown in Figure \ref{Figure 5}a, the PH with a bending angle $\beta$ = 12.1$^\circ$ is generated when the patchy cylinder has a rotation angle $\theta$ = $-$10$^\circ$. Increasing $\theta$ from $-$10$^\circ$ to 30$^\circ$ leads to the formation of PHs with a higher curvature (Figure \ref{Figure 5}d). The PH with a maximum bending angle $\beta$ = 23.4$^\circ$ can be generated at $\alpha$ = 30$^\circ$ (Figure \ref{Figure 5}b). Then, the curvature of the PHs becomes smaller as $\theta$ is further increased from 30$^\circ$ to 90$^\circ$ (Figure \ref{Figure 5}d). However, when $\theta$ is between 45$^\circ$ and 80$^\circ$, the light beams formed at the shadow side of the patchy cylinder is similar to a PJ, and its intensity distribution is approximately symmetric to the midline of the model, as shown in the inset of Figure \ref{Figure 5}d. The bending angle of the PH decreases to \mbox{$\beta$ = 15.2$^\circ$} at $\theta$ = 90$^\circ$ (Figure \ref{Figure 5}c).

{As shown by the black line in Figure \ref{Figure 5}e, the subtense L of the PH increases from \mbox{13.33 $\upmu$m} to 18.23 $\upmu$m when $\theta$ increases from $-$10$^\circ$ to 20$^\circ$, and then it decreases to 15.18 $\upmu$m when $\theta$ further increases to 45$^\circ$. When $\theta$ is between 45$^\circ$ and 80$^\circ$, the focused light has a structure similar to a symmetric PJ  (the inset of Figure \ref{Figure 5}d). The light beams show a curved shape again when $\theta$ is between 80$^\circ$ and 90$^\circ$ (Figure \ref{Figure 5}c), and the subtense L decreases from 17.68 $\upmu$m to 15.10 $\upmu$m when increasing $\theta$ from 80$^\circ$ to 90$^\circ$. The PH's height increment H at different rotation angles $\theta$ is shown as the red dots in Figure \ref{Figure 5}e, and its changing trend is similar to that of the subtense L. As shown by the red line in Figure \ref{Figure 5}e, the maximum height increment is obtained at $\theta$ = 30$^\circ$  with H = 1.56 $\upmu$m. Then, the height increment drops to H = 1.07 $\upmu$m when $\theta$ increases from 30$^\circ$ to 45$^\circ$. When $\theta$ is between 80$^\circ$ and 90$^\circ$, the height increment decreases as the rotation angle increases, with a value of H = 1.32 $\upmu$m ($\theta$ = 80$^\circ$) and H = 0.97 $\upmu$m ($\theta$ = 90$^\circ$), respectively.}

Figure \ref{Figure 5}f,g shows the value of I$_{max}$ (black line in Figure \ref{Figure 5}f), the corresponding full width at half maximum (FWHM) (red line in Figure \ref{Figure 5}f) and the position of I$_{max}$ (Figure \ref{Figure 5}g) at different rotation angles. The distance between the I$_{max}$ position and the right edge of the cylinder is denoted CP. We found that the patchy cylinder has the greatest focusing ability (Figure \ref{Figure 5}f) and the farthest focal point (Figure \ref{Figure 5}g) when $\theta$ is between 45$^\circ$ and 70$^\circ$, but the light beams focused by the patchy cylinder have a structure similar to a PJ, as shown in the inset of Figure \ref{Figure 5}d. Therefore, the patchy cylinder with a rotation angle \mbox{$\theta$ = 30$^\circ$} shows the strongest bending ability as well as a good focusing performance.

\begin{figure}[htb]
 
\includegraphics[width=0.8\textwidth]{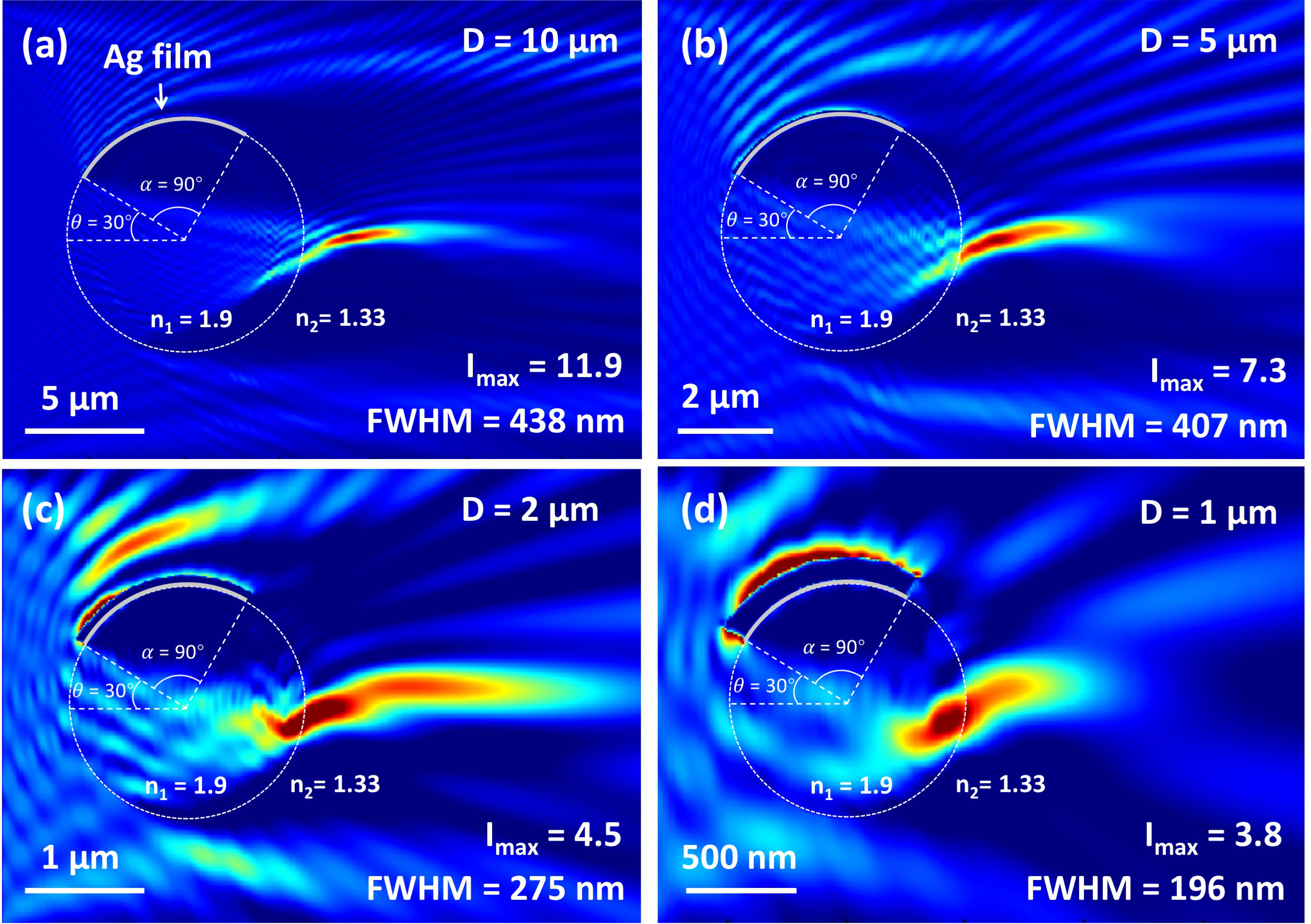}
\centering
\caption{Optical fields of the patchy microcylinders with $\alpha$ = 90$^\circ$ and $\theta$ = 30$^\circ$. The diameter of the cylinder is (\textbf{a}) 10 $\upmu$m, (\textbf{b}) 5 $\upmu$m, (\textbf{c}) 2 $\upmu$m and (\textbf{d}) 1 $\upmu$m, respectively.}
\label{Figure 6}
\end{figure}

Next, we changed the diameter of the patchy cylinders between 1 $\upmu$m and 10 $\upmu$m, while keeping the opening angle $\alpha$ of the Ag film as 90$^\circ$, the rotation angle $\theta$ of the patchy cylinder as 30$^\circ$, and the RI of the cylinder and the background medium as n$_1$ = 1.90 and n$_2$ = 1.33, respectively. As shown in Figure \ref{Figure 6}, in all four cases, the light beams with a curved structure are formed on the shadow side of the plane-wave illuminated patchy cylinder. Different from the results obtained using the patchy cylinders with half of their surfaces covered with Ag films ($\alpha$ = 180$^\circ$) (Figure \ref{Figure 4}e,f), the patchy cylinders with a smaller coverage of Ag film ($\alpha$ = 90$^\circ$) can generate PHs even when the diameter of the cylinder is below 5 $\upmu$m (Figure \ref{Figure 6}c,d). {The curvature $\beta$ of the PH increases and the subtense L and the height increment H of the PH decrease as the diameter of the cylinder decreases. The characteristics of the PHs are $\beta$ = 21.2$^\circ$, L = 6.31 $\upmu$m, H = 0.50 $\upmu$m (Figure \ref{Figure 6}a), $\beta$ = 24.3$^\circ$, \mbox{L = 3.26 $\upmu$m,} H = 0.21 $\upmu$m (Figure \ref{Figure 6}b), $\beta$ = 26.9$^\circ$, L = 2.40 $\upmu$m, H = 0.05 $\upmu$m (Figure \ref{Figure 6}c) and \mbox{$\beta$ = 28.4$^\circ$,} L = 0.55 $\upmu$m, H = 0.01 $\upmu$m (Figure \ref{Figure 6}d) for 10 $\upmu$m-, 5 $\upmu$m-, 2 $\upmu$m- and 1 $\upmu$m-diameter patchy cylinders, respectively.} We also found that the PHs generated from small particles have a smaller FWHM at the I$_{max}$ position. In this study, the minimum FWHM at the I$_{max}$ position is 196 nm, corresponding to $\sim$ 0.36 $\lambda$, which is generated from the 1 $\upmu$m-diameter patchy cylinder (Figure \ref{Figure 6}d).

{The patchy particles proposed in this work can be fabricated using the glancing angle deposition (GLAD) method.} The GLAD is a technique that transports vapor deposition to a target at an inclined angle relative to the plane of the particle arrays~\cite{08PawarAB, 09PawarAB}. In the GLAD process, particles within the same monolayer act as deposition masks for neighboring particles and thus patchy particles can be produced in a single deposition at a high yield of around 3.2 $\times$ 10$^5$ patchy particles per 1 cm$^2$ area for 2 $\upmu$m-diameter microspheres. The GLAD technique allows the precise positioning of patches onto the particle surface by controlling geometric parameters like the deposition angle, the diameter of particles, and so forth. In addition, the GLAD technique can be applied to the deposition of functional patches with various optical properties, such as the anti-reflection coating, thin-film polarizers, and so forth. Multiple patches can be fabricated on a single particle using GLAD technology multiple~times.

\section{Conclusions}

In conclusion, the generation of PHs from patchy microcylinders was investigated in detail in this study. The patchy microcylinders are dielectric cylinders whose surface is partially covered with Ag thin films. The fabrication of patchy cylinders can be realized using the GLAD method. Numerical simulation based on the FDTD method was used to investigate the characteristics of the PHs. {By rotating a 35 $\upmu$m-diameter patchy cylinder around its center, the bending angle of the PH can be changed between 12.1$^\circ$ and 23.7$^\circ$, the subtense of the PH can be changed between 13.33 $\upmu$m and 18.23 $\upmu$m, and the height increment of the PH varies between 0.67 $\upmu$m and 1.56 $\upmu$m.} PHs with a bending angle of 28.4$^\circ$ and a FWHM of 0.36 $\lambda$ can be obtained by using a 1 $\upmu$m-diameter patchy cylinder.

\printbibliography

\end{document}